\documentclass[final,sort&compress,numberedheadings]{aipproc}
\layoutstyle{8x11double}
\usepackage{graphicx}
\begin{document}

\title{Slow dynamics in cylindrically confined colloidal suspensions}
\classification{64.70.pv, 61.43.Fs, 82.70.Dd}
\keywords      {colloidal glass, confinement, glass transition}

\author{Nabiha Saklayen}{address={ Department of Physics, Emory
University, Atlanta, GA 30322 }}
\author{Gary L.~Hunter}{address={ Department of Physics, Emory
University, Atlanta, GA 30322 }}
\author{Kazem V.~Edmond}{address={ Department of Physics, Emory
University, Atlanta, GA 30322 }}
\author{Eric R.~Weeks}{address={ Department of Physics, Emory
University, Atlanta, GA 30322 }}

\date{\today}

\begin{abstract}
We study bidisperse colloidal suspensions confined within glass
microcapillary tubes to model the glass transition in confined
cylindrical geometries.  We use high speed three-dimensional
confocal microscopy to observe particle motions for a wide
range of volume fractions and tube radii.  Holding volume fraction
constant, we find that particles move slower in thinner tubes.
The tube walls induce a gradient in particle mobility:  particles
move substantially slower near the walls.  This suggests that the
confinement-induced glassiness may be due to an interfacial effect.
\end{abstract}

\maketitle

\section{Introduction}

Understanding the glass transition is one of the enduring
questions of solid-state
physics \cite{gotze92,ediger96,debenedetti01,cavagna09,dyre06,berthier11}.
The problem is simply stated:  in some cases, when a hot viscous
liquid is cooled, the viscosity rises dramatically but smoothly as
a function of temperature.  At some temperature the viscosity is
so large that the sample appears like a solid; this identifies the
glass transition temperature.  The same phenomenon can likewise
be induced by increasing the density (increasing the pressure)
\cite{roland05}.  In contrast to a regular phase transition
which occurs at well-defined temperatures and pressures, the
glass transition can depend on details such as the cooling rate.
Likewise, while phase transitions are signaled by abrupt changes
in the sample properties or their derivatives, the properties
of a glass-forming material such as viscosity and diffusivity
change smoothly, and to an extent the definition of the transition
temperature (or pressure) is a bit arbitrary
\cite{simionesco06,hunter12rpp}.
One of the key questions is what is changing microscopically
that is responsible for the macroscopic changes in viscosity; no
structural length scale has yet been found that would clearly explain
the viscosity change \cite{ernst91,charbonneau12}.

One clever way to probe length scales is to confine a sample:
rather than studying a macroscopically large sample (a ``bulk''
sample), studying a microscopic-scale sample.  Many experiments
show that glasses change their properties when their size is
sufficiently small, both for small-molecule glasses and polymer
glasses \cite{roth05,mckenna05,simionesco06,richert11}.  One of the
key observations is that the glass transition temperature $T_g$
changes for confined samples.  In some cases $T_g$ increases:
confined samples are glassier.  However, in other cases $T_g$
decreases.  The key difference explaining the increase or
decrease seems to be the boundary
conditions \cite{richert11}.  Samples with free surfaces, such as
thin free-standing polymer films, are less glassy (lower $T_g$).
Samples confined to pores or on substrates can be more glassy
(larger $T_g$), in particular if the sample molecules form strong
chemical bonds to the confining boundaries.

We wish to use colloidal samples as a glass-forming system
which can be studied in confinement.  Colloidal suspensions
are composed of solid particles in a liquid.  As the particle
concentration is increased, the sample becomes more and more
viscous \cite{marshall90,segre95,poon96,cheng02}.  Above a critical
concentration, the sample behaves as a glass, and a large number
of similarities have been observed between the colloidal glass
transition and glass transitions of polymers and small molecules
\cite{hunter12rpp}.  The most widely studied colloidal glass
transition is that of hard-sphere-like colloids, and the control
parameter is the volume fraction $\phi$ \cite{pusey86}.  The glass
transition point has been identified as $\phi_g \approx 0.58$, with
simulations demonstrating that this requires some polydispersity
\cite{zaccarelli09,pusey09}.

Experimentally, the colloidal glass transition
shifts to lower volume fractions in confined
samples:  confinement makes colloidal samples glassier
\cite{nugent07prl,edmond10b,edmond12,sarangapani08,sarangapani11,eral09}.
This has been studied exclusively in parallel-plate geometries,
where samples are confined between two glass walls that are
closely spaced.  Often these experiments use bidisperse samples
(mixtures of two particle sizes) so that the flat walls do not
induce crystallization \cite{nugent07prl,edmond12}.  An alternate
approach is to roughen the walls \cite{sarangapani08}.

The geometry of these prior colloidal experiments most closely resembles
thin films, which are used to study the glass transition of polymers
or small molecule glass formers in thin slits.  However, small
molecule glass formers are more commonly studied by using nanoporous
substrates; a variety of these nanoporous substrates are reviewed
in Ref.~\cite{simionesco06}.  Some of these substrates are quite
disordered with pores of a variety of shapes and sizes.  Others are
ordered: for example, porous oxide ceramics have a regular
lattice of cylindrical nanopores with well defined sizes
\cite{morineau02,simionesco06}, as do anodized aluminum oxide
membranes \cite{serghei10b}.  Experiments find that confinement
in cylindrical pores can both enhance or diminish glassy behavior
\cite{morineau02}.  Simulations show that the boundaries play
an important role in this:  rough walls that frustrate layering
of particles result in glassier dynamics, and smooth walls result
in less glassy dynamics \cite{kob00,scheidler02}.

In this paper, we present a study of colloidal samples confined
in cylindrical glass tubes, to mimic the geometry of cylindrical
nanopores.  We use confocal microscopy to observe both the structure
and dynamics of the samples.  Similar to prior colloidal work, we
find that confined colloidal samples are glassier.  In particular,
particle motion is dramatically slower at the capillary tube walls,
demonstrating that we see an interfacial effect.  We use a
bidisperse
sample to prevent confinement-induced crystallization or other
ordering, known to occur for monodisperse cylindrically confined
spheres \cite{harris80,pittet96,moon04,li05,tymczenko08,lohr10}.
Nonetheless, the particles layer against the walls, and these
layers slightly influence the motion in ways similar to previous
observations \cite{nugent07prl,edmond12}.  The data add to the
analogy with confined small-molecule glass-formers.  Additionally,
they are of interest for colloidal suspensions themselves:  the
implication is that for microfluidic applications, it will be more
difficult than anticipated to flow dense colloidal suspensions, as
they will be glassier in small tubes than an equivalent bulk sample.
Colloids confined in cylinders or small channels have been
studied before, but only in dilute concentrations \cite{eral10}
and/or in extremely thin channels that are only one particle diameter
across \cite{valley07,wonder11}.

\section{Experimental methods}
\label{methods}

Our goal is to use hard-sphere-like colloids as a model system.
Colloids have proved to be effective models with similarities
to hard-sphere computer simulations; see \cite{hunter12rpp} for
a discussion.  A hard-sphere system means that the particles
do not interact with one another beyond their radius and are
infinitely repulsive at contact \cite{alder57,woodcock81,speedy98}.
An advantage is that the particle size can be selected to be $\sim
1$~$\mu$m in radius:  small enough to undergo random Brownian
motion, yet still large enough to be imaged using microscopy
\cite{hunter12rpp}.

We use poly(methyl methacrylate) (PMMA) spheres coated with a
polymer brush layer that sterically stabilizes the particles,
preventing them from aggregating \cite{antl86}.  We use a
bidisperse
mixture with particles of two different radii, large particles
with radius $a_L=1.08$~$\mu$m and small particles with radius
$a_S=0.532$~$\mu$m, helping us avoid crystallization \cite{lohr10}.
The particles have a polydispersity of approximately 6\% and
additionally their mean radii $a_L$ and $a_S$ are each uncertain
by 1\%.  Our particles are fluorescently dyed so that we can observe
their motion with confocal microscopy.  It is desirable to reduce
the influence of gravity in a colloidal suspension by density
matching the colloid with the surrounding fluid.  To do this we
use a standard mixture of 85\% (weight) cyclohexyl bromide and
15\% decahydronaphthalene (decalin, mixture of cis- and trans-).
This solvent mixture also matches the particles' refractive index,
which is necessary for the microscopy.  To reduce the influence of
electrostatic repulsion between the particles, we saturate the
solvent mixture with tetrabutylammonium bromide (Aldrich, 98\%)
with a resulting concentration of $\sim 190$~$\mu$M
\cite{yethiraj03}.

\begin{figure}[bhtp]
\centering
\includegraphics[scale=0.40]{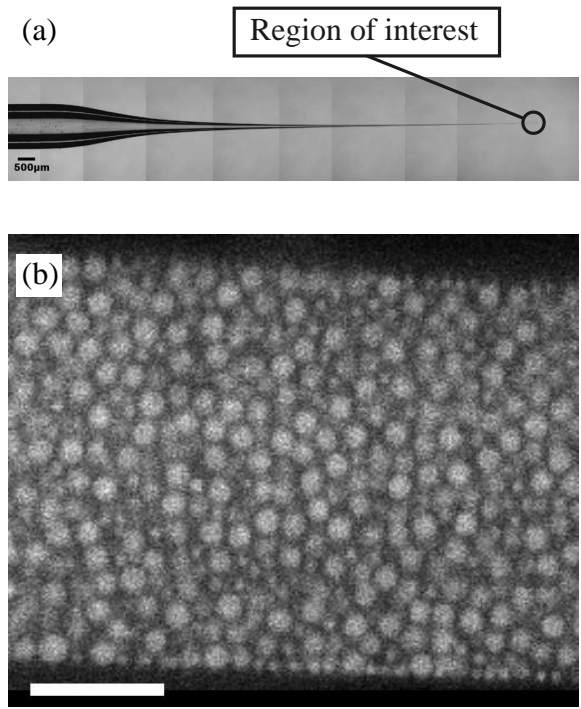}
\caption 
{
Top:  Composite photograph of capillary tube.  The scale bar is
500~$\mu$m.  Bottom:  2D confocal image from experiment 55,
with volume fraction $\phi_{\rm tot}=0.49$.
The scale bar is 10~$\mu$m.
\label{setup}
} 
\end{figure}

Our sample chambers are glass capillary tubes as shown in
Fig.~\ref{setup}(top).  The capillary tip was made using an
automated pipette puller. To help the capillary tube fit onto a
microscope slide, the large end of the capillary tip was
cut off while ensuring not to break the thin end of the capillary
tube.  The thin end was often too thin, so it too was cut,
leaving an opening a few microns in diameter.
The shortened capillary tube was then dipped into a vial
containing the colloidal suspension for 10 seconds or so, and the
sample flows into the glass tube due to capillary forces.
We use quick drying UV epoxy (Norland 81)
to seal both ends of the tube and also to glue the tube to the slide.
In fact, it is useful to cover the
tip entirely with glue to ensure stability.  The microscopy is not
affected too much, as the sample and epoxy have similar indices
of refraction (1.495 and 1.56 respectively).  The result is a
confined sample, such as shown in Fig~\ref{setup}(bottom), which
can be studied at different locations to give different tube sizes.
The capillary tubes vary slightly in radius as a function
of length (slope of about $1^\circ$), and the slight taper does
not seem to affect our overall results.  The taper is slight
enough that we cannot observe it in any of our confocal images.
Due to the pipette puller protocol we used, the capillary tubes
sag slightly, and so their cross-section is slightly elliptical
rather than circular.  This does not seem to influence our results
and will be discussed further below.  When the tubes are filled
with the sample, some of the particles stick irreversibly to the
tube walls due to van der Waals forces.  This typically resulted
in one complete layer of particles (of both sizes) coating the
wall.  During the course of our experiments, we do not observe
any new particle sticking to the walls, nor do we observe any
stuck particles becoming unstuck.

We use a confocal microscope (Visitech vt-Eye) to study our
samples.  In a confocal microscope, laser light is scanned across
the fluorescent sample and excites the dye to emit a different
color light.  The emitted light passes through a pinhole to
remove the out-of-focus light and then is measured by a detector,
a photomultiplier tube in our microscope.  The sample is quickly
scanned in $x$ and $y$ to acquire a two-dimensional (2D) image.
Then the microscope focus is adjusted to scan more 2D images at
different depths $z$, thus building up a 3D image.  Being able
to create 3D images makes confocal microscopy a powerful tool
for studying particle dynamics in the system.  Each 3D scan takes
about 2-3 seconds, and we typically take movies comprised of 400 of
these 3D images.  The particles are tracked in 3D using standard
tracking techniques \cite{crocker96,dinsmore01}.  Within each
confocal image, the small and large particles are easy to tell
apart [see Fig.~\ref{setup}(bottom)] and so we can distinguish
between them in our data \cite{narumi11}.

We have two control parameters, the tube radius and the volume
fraction.  Complete descriptions of how these are measured are
given in the Appendix; here, we summarize the key points.

Our first control parameter is the tube radius.  The positions of
the particles give us an accurate idea where the tube surface is.
However, the tubes have an elliptical cross-section rather than a
circular cross section.  We measure the major and minor axes $R_{\rm
max}$ and $R_{\rm min}$ for each experiment.  The ratio $R_{\rm
max}/R_{\rm min}$ ranges from 1.14 to 1.39, with mean 1.24.  Because
we are concerned with confinement effects, we report our data in
terms of $R_{\rm min}$ in general, although both radii are listed
for all experiments in Table I.  Note that we report the radii
corresponding to the maximum positions of the observed particle
centers:  in general the particles at these maximum positions are
the small ones (whose centers can get closer to the tube walls)
and so the true tube sizes are larger by $a_S=0.532$~$\mu$m.

Our other key control parameter is the volume fraction $\phi$.
We measure this in each data set by counting the numbers of
small and large particles observed within a subvolume of the
tube, and converting this to the volume fraction using the known
particle sizes.  Note that 1\% uncertainties in the particle radii
translate to 3\% uncertainties of the volume fraction, and since
each particle radius is uncertain, we have an overall systematic
volume fraction uncertainty of at least 5\% \cite{poon12}.

In summary, our experimental method is to image different portions
of the same tube in hopes to get a constant volume fraction with
differing tube radii, and to study different tubes with different
volume fractions to understand the role of volume fraction.
In practice, we determine these variables when the data are
post-processed, and report the measured values in Table I.

\begin{table}
\begin{tabular}{cccccccc}
Expt & $\phi_{\rm tot}$ & $R_{\rm min}$ & $R_{\rm max}$ & $\langle \Delta
x^2 \rangle$ & $N_{\rm small}/N_{\rm tot}$ \\
\hline
9b  &   0.19 & 11.1 & 14.9 & 4.8  & 0.45 \\ 
9a  &   0.19 & 12.7 & 17.4 & 4.7  & 0.51 \\ 
5a  &   0.20 & 10.5 & 14.2 & 2.9  & 0.37 \\ 
12b &   0.22 & 10.7 & 14.9 & 3.7  & 0.44  \\ 
11b &   0.22 &  8.5 & 11.3 & 2.2  & 0.37  \\ 
63  &   0.43 &  6.5 &  7.4 & 0.43 & 0.43 \\
54  &   0.45 & 10.9 & 13.2 & 0.86 & 0.33 \\
52  &   0.46 & 13.1 & 15.9 & 1.0  & 0.18 \\
55  &   0.49 & 12.8 & 15.4 & 0.91 & 0.18 \\
56  &   0.49 & 14.3 & 17.2 & 0.70 & 0.14 \\
51  &   0.50 & 17.1 & 20.7 & 0.99 & 0.07 \\
62  &   0.51 &  8.6 &  9.8 & 0.29 & 0.17 \\
65  &   0.53 &  8.8 & 10.1 & 0.22 & 0.16 \\
64  &   0.54 &  7.8 &  8.9 & 0.11 & 0.22 \\
\hline
\end{tabular}
\caption{
List of experiments, ordered by total volume fraction $\phi_{\rm
tot}$.  The volume fraction of the small particles can
be determined by $\phi_S = \phi_{\rm tot} [1 + ((1-f)/f)(a_L/a_S)^3]^{-1}$,
where $f \equiv N_{\rm small}/N_{\rm tot}$.  The volume fraction of
the large particles is then $\phi_L = \phi_{\rm tot} - \phi_S$.
The tube sizes $R_{\rm min}$ and $R_{\rm max}$ correspond to the
maximum radii that the particle centers can reach; the physical
tube walls are a distance $\approx a_S$ further away.
The value of $\langle \Delta x^2 \rangle$ given is for $\Delta t =
100$~s, and corresponds to the information plotted in
Figs.~\ref{msds}, \ref{msdphase}.
}
\label{table1}
\end{table}

\section{Results}

\subsection{Motion slows in confined samples}
\label{resultsgeneral}

\begin{figure}[bhtp]
\centering
\includegraphics[scale=0.50]{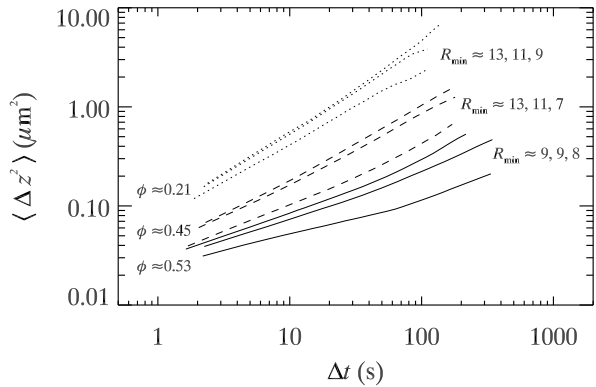}
\caption 
{
\label{msds}
Mean square displacements in the $z$ direction (along the cylinder
axis).  Each family of curves is labeled at left by the volume
fraction, and at right by the values of the minimum cylinder radius
(in $\mu$m).
For each curve, the volume fraction is within $0.02$ of the labeled
value.
} 
\end{figure}

\begin{figure}[bhtp]
\centering
\includegraphics[scale=0.50]{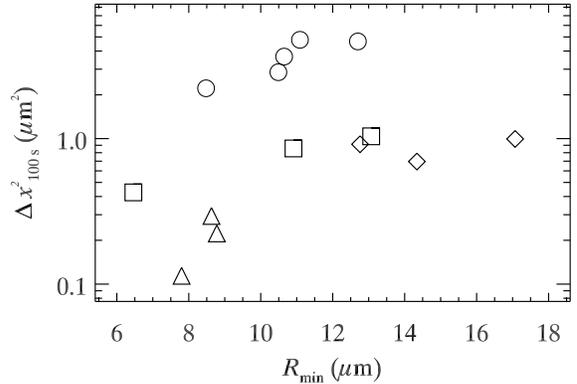}
\caption 
{
The value of $\langle \Delta x^2 \rangle$ at the time scale
$\Delta t = 100$~s,
plotted as a function of the cylinder minimum radius $R_{\rm min}$.
The symbols indicate the volume fraction:  
circles are $\phi = 0.21 \pm 0.02$, 
squares are $\phi = 0.45 \pm 0.02$,
diamonds are $\phi = 0.49 \pm 0.01$, and
triangles are $\phi = 0.53 \pm 0.02$.
The circles correspond to the dotted lines in Fig.~\ref{msds}, the
squares correspond to the dashed lines, and the triangles correspond
to the solid lines.
See Table I for a full listing of all volume fractions, mobility
values, and other details.
} 
\label{msdphase}
\end{figure}

By following the motion of all of the particles in 3D, we can observe
how the motion depends on confinement.  We quantify this by calculating
the mean square displacement, defined as
\begin{equation}
\langle \Delta z^2 \rangle = \langle [z(t+\Delta t) -
z(t)]^2 \rangle
\end{equation}
where $\langle \Delta z^2 \rangle$ is a function of the lag
time $\Delta t$, and the angle brackets indicate an average
over all times $t$ and all particles.  Here we are considering
the $z$ direction to be along the axis of the tube, primarily
because this axis is perpendicular to the optical axis of the
microscope and therefore has less position uncertainty\footnote{The mean
square displacement for the other components are qualitatively
similar, except that they have more uncertainty which
artificially increases the data at small lag times; see
\cite{poon12} for a discussion.}. The data are plotted in
Fig.~\ref{msds}, where each family of curves correspond to a
different volume fraction.  Within each family, the slower curves
correspond to narrower tubes: confinement induces glassier behavior.
To further quantify this, we consider the specific value of $\langle
\Delta x^2 \rangle$ at a lag time $\Delta t = 100$~s (which is
arbitrary but chosen to match prior work \cite{nugent07prl}).
This value is plotted as a function of minimum tube radius
$R_{\rm min}$ in Fig.~\ref{msdphase}.  The different symbols
correspond to different ranges of volume fractions, and in general
within each range, smaller tubes correspond to slower motion.
Plotting the data against $R_{\rm max}$, $(R_{\min}+R_{\max})/2$,
or $\sqrt{R_{\rm min}R_{\rm max}}$ does not change the overall
appearance of this graph significantly.  It is intriguing to note
that the magnitude of the effect is less significant than was seen
in similar experiments with parallel plates \cite{nugent07prl}.
Here, the strongest influence of confinement is to lower mobility
by a factor of $\sim 3$ (for circles, squares, and triangles
in Fig.~\ref{msdphase}).  With parallel plates, mobility was
lowered by a factor of $\sim 40$ for data with $\phi \approx 0.46$
\cite{nugent07prl}, a volume fraction in the same range as our
cylindrical data.  We are unsure why the two experiments differ
in the magnitude of mobility reduction.

\subsection{Motion is slower near walls}

\begin{figure}[bhtp]
\centering
\includegraphics[scale=0.90]{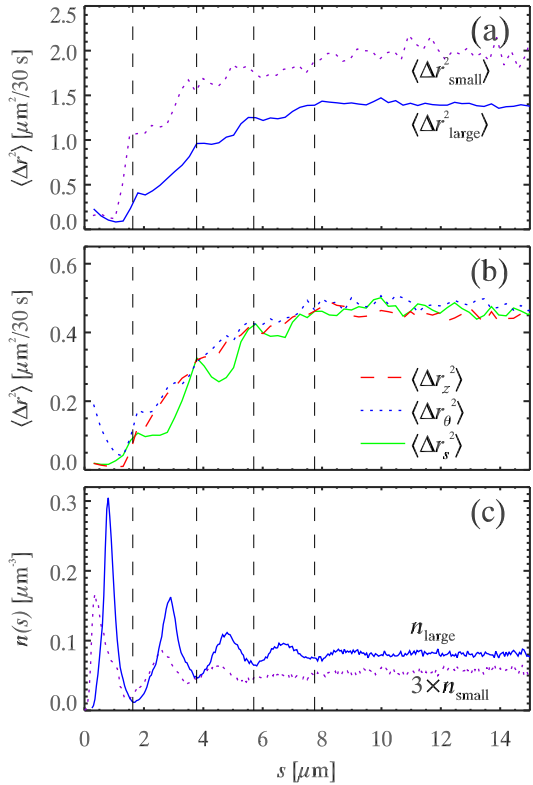}
\caption 
{
(Color online)
(a) Mobility as a function of distance $s$ from the wall, for the
small particles and large particles as indicated.  (b) Mobility
for the large particles only, for the components of motion as
indicated.  (c) Number densities of the small and large
particles.  Note that the number density of the small particles
has been multiplied by four.  In all panels, the vertical dashed
lines correspond to the local minima of $n_{\rm large}$.  They
are spaced approximately 2.04~$\mu$m apart.  The data are from
experiment 51 with $\phi_{\rm tot}=0.50$ and
$R_{\rm min}=17.1$~$\mu$m; see Table I for other details.
\label{mobilitybig}
} 
\end{figure}

Microscopy allows us to spatially resolve details of the motion.
If confinement-induced slowing is a finite size effect, then
the motion might be spatially homogeneous:  the whole sample
feels that it is small \cite{he07,richert11}.  If instead the
confinement-induced slowing is an interfacial effect (due to
the sample-wall interface), then particle motion would depend
on where each particle is relative to the boundary.  Of course,
both effects could be present simultaneously \cite{morineau02}.
We check this by plotting the particle mobility $\langle \Delta
r^2 \rangle$ as a function of the distance $s$ to the nearest
wall in Fig.~\ref{mobilitybig}(a).  It is immediately apparent
that particles move slower when they are close to the wall
($s \rightarrow 0$), and a plateau value for the mobility isn't
reached until several particle radii into the sample.  Note that
we are using the full 3D mobility $\langle \Delta r^2 \rangle =
\langle \Delta x^2 + \Delta y^2 + \Delta z^2 \rangle$, for a fixed
lag time $\Delta t = 30$~s (for which we have more statistics), and
now the angle brackets indicate an average over those particles with
the specific value of $s$.  The value of $s$ is based on the initial
position of the particle, at time $t$ rather than $t+\Delta t$.
It is probable that the mobility near the wall is very slightly
enhanced, due to particles which diffuse away from the wall
during $\Delta t$ and thus enhance their mobility \cite{eral10}.
The different curves are for small and large particles as indicated.
Not surprisingly, the smaller particles are more mobile than
the large ones.  Our directly observed gradient in mobility is
similar to that inferred from experiments on small molecule glasses
\cite{he05,he07,lequellec07}, polymer glasses \cite{ellison03},
and seen directly in simulations \cite{kob00,scheidler02,teboul02}.
Our data show that the mobility changes over a distance of several
particle diameters.

A second feature of the data of Fig.~\ref{mobilitybig}(a) is that
the curves oscillate.  The oscillations are related to the
fluctuations of the particle number density, as seen by comparing
Fig.~\ref{mobilitybig}(a) to Fig.~\ref{mobilitybig}(c).  The
latter shows the fluctuations of the number density of large
and small particles (solid and dotted lines, respectively).  The
mobility data in  Fig.~\ref{mobilitybig}(a) are anticorrelated
with $n_{\rm large}(s)$:  the local minima of $n_{\rm large}(s)$
are indicated by the vertical dashed lines, and
correspond to local maxima of $\langle \Delta r^2 \rangle$.  The
influence of $n_{\rm small}$ is not seen, probably because the
number fraction of small particles is much less than that of
the large particles ($N_{\rm small}/N_{\rm large} = 0.07$ for
these data).  The anticorrelation between number density and
mobility matches what has been seen in prior work on confined
samples \cite{nugent07prl,edmond12}.

Further insight into the mobility is found by splitting $\langle
\Delta r^2 \rangle$ into components parallel to the tube ($z$
direction), tangential to the tube wall ($\theta$ direction), and
in the radial direction ($s$ direction).  These components of
mobility are shown in Fig.~\ref{mobilitybig}(b).  Here it is
apparent that the radial mobility is the most influenced by the
oscillations of $n_{\rm large}(s)$.  Again, this is the same
result seen before with flat walls \cite{nugent07prl}.  The
particles at the maxima of $n_{\rm large}(s)$ appear to be at
favorable positions and their mobility is reduced, whereas those
at the minima are in less favorable positions with higher
mobility.  Motions in the $z$ and $\theta$ directions are along
contours of constant mean $n$ and do not fluctuate with $n(s)$.

\begin{figure}[bhtp]
\centering
\includegraphics[scale=0.90]{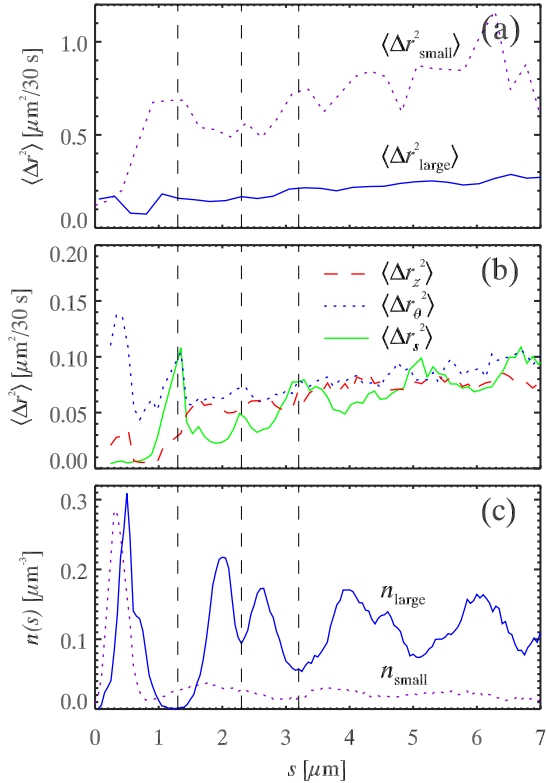}
\caption 
{
(Color online)
(a) Mobility as a function of distance $s$ from the wall, for the
small particles and large particles as indicated.  (b) Mobility
for the large particles only, for the components of motion as
indicated.  (c) Number densities of the small and large
particles.
In all panels, the vertical dashed
lines correspond to some of the local minima of $n_{\rm large}$.
The data are from
experiment 64 with $\phi_{\rm tot}=0.54$ and
$R_{\rm min}=7.8$~$\mu$m; see Table I for other details.
\label{mobilitytiny}
} 
\end{figure}

All of the results of Fig.~\ref{mobilitybig} are qualitatively
replicated in Fig.~\ref{mobilitytiny}, which is data from a
smaller radius tube ($R_{\min}=7.8$~$\mu$m compared to $R_{\rm
min}=17.1$~$\mu$m) and volume fraction only slightly larger than
that of Fig.~\ref{mobilitybig} ($\phi_{\rm tot}=0.54$ compared to
0.50).  In Fig.~\ref{mobilitytiny}(a), the mobility is lower near
the boundary, lower for large particles, and oscillates with
higher mobility corresponding to minima of $n_{\rm large}(s)$.
The data for the components [Fig.~\ref{mobilitytiny}(b)] are
noisier due to less statistics in the smaller tube, but again
$\langle \Delta r_s^2 \rangle$ shows a stronger anticorrelation
with $n_{\rm large}(s)$.  The oscillations of $n_{\rm large}(s)$
in Fig.~\ref{mobilitytiny}(c) are more complex than those seen in
Fig.~\ref{mobilitybig}(c), probably due to packing constraints of
a smaller tube.

\section{Conclusions}

Our results -- in particular Figs.~\ref{mobilitybig}(a)
and \ref{mobilitytiny}(a) -- suggest that the slower motion
of confined colloidal samples is due to an interfacial effect,
where particles near the sample walls are slowed.  This agrees
with a prior observation that colloidal particle motion is slower
near rougher walls \cite{edmond10b}, a result demonstrating that
the nature of the confining walls plays a role and not merely the
finite size of the sample chamber.  It is unlikely this is merely
a hydrodynamic effect, as the magnitude of such an effect would
only be a factor of $\sim 2-4$ in mobility and would not depend on
$R_{\rm min}$
\cite{eral10,edmond12}.

The clear gradient seen in Figs.~\ref{mobilitybig}(a) and
\ref{mobilitytiny}(a) was not seen in prior work where colloids
were confined between parallel walls \cite{nugent07prl}.  It may be
that the influence of confinement is stronger in the cylindrically
confined case:  between parallel walls, there are two unconfined
directions, whereas in the cylinder there is only one unconfined
direction \cite{barut98}.  Another possibility is that the curvature of the
cylindrical walls introduces an effect not present with flat
walls\footnote{Although the hydrodynamic effect of the wall should
be the same for flat or curved walls, sufficiently close to the
walls.  Prior simulations found essentially no difference in
the diffusion constant between flat and curved walls, for
particles within 1.5 diameters of the wall \cite{eral10}.}.
These geometrical differences are the only major differences
between the experiments in this paper and those published earlier.
As noted in Sec.~\ref{resultsgeneral}, the other observed
difference is that the slowing of motion in cylindrically confined
samples (Fig.~\ref{msdphase}) is less pronounced compared to
the observations between parallel plates \cite{nugent07prl}.
This is counterintuitive given the argument above that cylindrical
confinement is ``stronger'' than parallel-plate confinement.
This trend is the opposite of that seen for small molecule liquids
\cite{barut98}.  Future experiments may be able to elaborate on
the question for colloids:  at one extreme, particle motion could
be studied in a half-infinite system near a wall, whereas at the
other extreme, particle motion could be studied in a spherical
pore.  The data presented in this paper suggest that changing
the dimensionality of the confinement in this way can result in
interesting and qualitatively distinct behavior, in other words,
enhancing a mobility gradient near walls while diminishing the
overall confinement effect.

Our results also imply that flowing colloidal suspensions through
small cylindrical tubes will be harder for smaller tube radii.
Additionally, the decrease in mobility near the walls perhaps
implies an increase in the apparent viscosity of the sample
near the walls, thus modifying the flow velocity profile in a
nontrivial fashion.

Note that we do not see any quantization effects:  we do not
see any particular change in the dynamics at any special ratios
of particle sizes to tube sizes.  This is in contrast to some
theoretical predictions \cite{mittal08,desmond09,lang10}.  However,
our data are only at a limited number of tube sizes, as shown in
Fig.~\ref{msdphase}; our tubes are elliptical in cross section and
so the ratio of particle size to tube size is not a constant for
any given data set; and it is likely that such quantization effects
are more subtle than we would be able to see in an experiment.

\begin{theacknowledgments}
Funding for this work was provided by the National Science
Foundation (Grant No.~DMR-0804174) and by an Emory University
SIRE grant to N.~S.  We thank C.~B.~Roth for helpful discussions.
\end{theacknowledgments}

\section*{Appendix}

We explain in more detail how our experimental parameters are
measured.  In each case, we prepare samples and take data,
measuring the tube radii and volume fraction from the data.

{\bf Tube radius:}
As noted in Sec.~\ref{methods}, the tubes do not typically have a
circular cross-section.  Also, in general, the images of the tube
are not precisely aligned with the $xyz$ laboratory reference frame.
To determine the radius, the position data are first rotated by
$0-4^\circ$ around the $x$ and $y$ axes as necessary so that the
$z$ axis of the data corresponds with the tube axis.  (Note that
the $z$ axis is {\it not} the optical axis of the microscope;
rather, the $z$ axis is within a few degrees of perpendicular
to the optical axis.)  Then the data are projected onto the $xy$
plane and their center of mass is found.  The $x,y$ coordinates
are converted to $r,\theta$ and the maximum $r$ is found as a
function of $\theta$.  Following the procedure of Eral {\it et
al.}, in order to smooth the measured contour we fit $r(\theta)$
to a Fourier series up to $m=4$ modes \cite{eral10}.  The function
$r(\theta)$ is found to be well-described by an ellipse in all
cases, and accordingly it is easy to determine the major and
minor axes.

{\bf Volume fraction:}
For our experiments, we take movies from different portions of
several tubes.  Ideally each tube is filled with a sample of
homogeneous volume fraction, but in practice the volume fraction
varies slightly from region to region.  Additionally, defining
volume fractions in confined sample chambers is a little problematic
as the concentration is inherently smaller at the walls simply due
to packing constraints \cite{desmond09} even if the interparticle
spacing is spatially homogeneous.  To define volume fraction,
we integrate the $r(\theta)$ data described in the previous
paragraph to determine the cross sectional area (adding on $a_S$ to
determine the physical wall boundary).  The length of the observed
region is known, so therefore we know the volume $V_{\rm tot}$
of the tube that is imaged.  Likewise we know the numbers of the
small and large particles $N_S$ and $N_L$ that are in the image,
and so the volume fraction can be determined from $(N_S V_S +
N_L V_L)/V_{\rm tot}$ in terms of the individual particle volumes
$V_S$ and $V_L$.  $V_S \sim a_s^3$ and likewise for $V_L$, so 1\%
uncertainties in the particle radii lead to 3\% uncertainties of
the volume fraction.  Since each particle radius is uncertain,
we have an overall systematic volume fraction uncertainty of at
least 5\% \cite{poon12}.  There is also some uncertainty between
samples as the different samples are observed to have different
number ratios of small and large particles (see Table I), and so
errors in small and large particle radii will affect the different
volume fraction calculations in different amounts.  Encouragingly,
visual inspection of the images suggests that the calculated
volume fractions listed in Table I are at least qualitatively in
correct order.  Samples with volume fractions within 0.02 of each
other appear visually to be the same volume fraction, and samples
with greater differences are visually distinct.

\bibliographystyle{aipproc}   
\bibliography{eric}

\end{document}